\documentstyle[aps,prd,preprint]{revtex}

\def\r{\rho}

\tighten
\begin{document}
\draft

\title{Cauchy-characteristic Evolution of Einstein-Klein-Gordon Systems}

\author{Roberto G\'{o}mez,${}^{(1)}$ Pablo Laguna,${}^{(2)}$ 
Philippos Papadopoulos${}^{(2)}$ and Jeff Winicour${}^{(1)}$}

\medskip

\address{
${}^{(1)}$ Department of Physics \& Astronomy\\
University of Pittsburgh, Pittsburgh, PA 15260
}

\medskip

\address{
${}^{(2)}$ Department of Astronomy \& Astrophysics and \\
Center for Gravitational Physics \& Geometry\\
Penn State University, University Park, PA 16802
}

\maketitle

\begin{abstract}

A Cauchy-characteristic initial value problem for the
Einstein-Klein-Gordon system with spherical symmetry is presented.
Initial data are specified on the union of a space-like and null
hypersurface.  The development of the data is obtained with the
combination of a constrained Cauchy evolution in the interior domain
and a characteristic evolution in the exterior, asymptotically flat
region.  The matching interface between the space-like and
characteristic foliations is constructed by imposing continuity
conditions on metric, extrinsic curvature and scalar field variables,
ensuring smoothness across the matching surface. The accuracy of the
method is established for all ranges of $M/R$, most notably,
with a detailed comparison of invariant observables against 
reference solutions obtained with a calibrated, global, null algorithm.

\end{abstract}

\widetext

\section{Introduction}
\label{sec:introduction}

The correct physical formulation of any asymptotically flat, radiative
Cauchy problem requires boundary conditions at spatial infinity. These
conditions ensure not only that total energy and the energy loss by
radiation are both finite, but they are also responsible for the proper
$1/r$ asymptotic falloff of the radiation fields.  However, when
treating radiative systems computationally, an outer boundary must be
established artificially at some large but finite distance in the wave
zone, i.e. many wavelengths from the source. Imposing an accurate
radiation boundary condition at a finite distance is a difficult task
even in the case of simple radiative systems evolving on a fixed
geometric background.  The problem is exacerbated when dealing
with the Einstein equations.

In recent years, the characteristic initial value problem (cIVP)
formulation of the Einstein equations in non-spherically symmetric
configurations \cite{Isaacson,Null-cone} has been explored, providing
possible alternatives to the practical and theoretical problems
introduced by the outer boundary conditions of the Cauchy initial value
problem (CIVP).  Based on the concept of combined Cauchy-characteristic
evolution, a number of systems are currently under investigation.  The
motivation behind these new formulations of the initial value problem
is to capture the advantages that each approach exhibits and at the
same time avoid some of the corresponding drawbacks. The cIVP permits
the construction of integration algorithms that allow null infinity to
be included in a compactified grid, hence facilitating and clarifying
the extraction of radiation. However, characteristic formulations
generally break down in regions of complicated caustic structure, which
are unavoidable in strongly asymmetric geometries such as those
describing the merger of two black holes. Cauchy evolutions avoid this
problem, yet they provide no natural way to impose conditions at the
outer boundary.  Since each method operates successfully precisely in
the region where the other has its shortcomings, an appropriate
matching of the two initial value formulations promises an effective
approach to the outer boundary condition and the caustic problem.

The general idea behind the Cauchy-characteristic matching (CCM) is not
entirely new. An early mathematical investigation exhibiting unions of
space-like and characteristic surfaces was given in \cite{Duff}.
Regarding general relativistic systems, a discussion of the potential
of the method appears in \cite{Bishop}. The concept of a null exterior
attached to a Cauchy evolution appears also in connection with
perturbative approaches to the outer boundary problem, e.g., in
\cite{Seidel} as well as in \cite{Hobill}. Yet, only recently
\cite{ClarkeI,ClarkeII,ClarkeIII,3Dccm}, has the concept been carefully
explored with respect to its practicability.  A detailed study of the
stability and accuracy of CCM for linear and non-linear wave equations
has been presented in \cite{3Dccm,3Dccmlong}, illustrating its
potential for a wide range of wave systems.  The numerical
investigation of cylindrically symmetric solutions of the Einstein
equations has also been carried out \cite{ClarkeII,ClarkeIII}.

The objective of this paper is to develop and carefully calibrate the CCM
method for asymptotically flat, spherically symmetric space-times,
which are evolving in the presence of a minimally coupled,
self-gravitating, massless scalar field. This is an initial step towards
developing a general method applicable to the full Einstein equations.
Research on this topic is stimulated and guided by the requirements of
the Binary Black Hole Grand Challenge Alliance, a major collaboration
aimed at the investigation of the merger of two in-spiraling black
holes. The CCM approach will provide, in this
context, both boundary conditions and radiation waveform
extraction. The model problem investigated herein captures some
essential aspects of the general system, including wave
propagation on dynamical backgrounds and black-hole formation.  The
dynamics of the model is governed by the coupled Einstein-Klein-Gordon
equations in spherical symmetry,
\begin{equation}
    G_{ab}=\kappa (\nabla_a \Phi\nabla_b \Phi + L g_{ab})~,
\label{eq:ein}
\end{equation}
and 
\begin{equation}
    \nabla_a\nabla^a \Phi = 0,
\label{eq:klein}
\end{equation}
where $\Phi$ is a scalar field whose
Lagrangian is
\begin{equation}
      L = -{1\over 2}\nabla_a \Phi\nabla^a \Phi
\label{eq:lagran}
\end{equation}
and $\kappa=8\pi$. (Units are chosen so that $G=c=1$). 
Space-time indices are denoted with Latin letters ($a,b,c,\dots$) 
and space indices with ($i,j,k,\dots$).

This system has been studied intensively in the last five years both
analytically \cite{Christo} and numerically
\cite{Asymptotics,Gundlach,Choptuik-PRL,Goldwirth,Choptuik-comp}.  
The existent understanding of the system is quite appropriate, as
the exploration and calibration of the combined
evolution is done in a non-trivial geometric setting, yet with good
basic knowledge of the expected physical behavior.

The interior foliation and the associated integration algorithm are
presented in Section~\ref{sec:cauchy}. The time integration is
analogous to that used by Choptuik~\cite{Choptuik-print}, the main
difference being a different choice of boundary conditions.
Section~\ref{sec:null} describes the characteristic evolution
algorithm.  The method closely follows~\cite{Asymptotics}, with minor
changes to accommodate the matching interface. In Section~\ref{sec:ccm}
the geometric concepts underlying the matching theory are discussed,
and a number of simplifying assumptions are put forward, which lead to
a simple set of matching conditions for the problem at hand.  The
implementation of the combined evolution, along with validation tests,
and numerical experiments covering a wide range of methodologies and physical
parameters are given in Section~\ref{sec:tests}.

\section{Interior Domain: Cauchy Evolution}
\label{sec:cauchy}

For the interior (Cauchy) domain $M^-$ of the four-dimensional
space-time, Einstein's equations will be written using a standard
3$+$1 ADM decomposition. Generally, a foliation 
of space-like slices (hypersurfaces) is constructed and labeled
by a scalar function $\tau$. The unit normal to these slices is
$n^a$. By construction, $n_a = -\alpha \nabla_a \tau$, where $\alpha$
is the lapse function.  The intrinsic metric on each slice is then given by $
\gamma_{ab} = g_{ab} + n_an_b$.  The vector $\alpha n^a$ connects the
slices of the foliation; however, this time vector is not unique. In
general, the vector $t^a$ can be chosen as $t^a = \alpha n^a +
\beta^a$ where $\beta^a$ is the shift vector and $\beta^a n_a = 0$.
Adopting this decomposition, the four dimensional metric element in
the case of spherically symmetric space-times reduces to the
3$+$1 form 
\begin{equation}
   ds^2 = (-\alpha^2 + \beta^2 e^{-2A}) dt^2 + 2  \beta\,dt\,d\r  + 
   e^{2A} \,d\r^2 + \r^2 e^{2B} \, (d\theta^2 + \sin{^2\theta}\,d\phi^2),
\label{eq:met} 
\end{equation}
where $\beta \equiv \beta_\r$ is the only non-zero component of the
shift vector, and the metric on the space-like hypersurfaces takes the
form $\gamma_{ij} = \text{diag}(e^{2A}, \r^2 e^{2B} , \r^2 e^{2B}
\sin{^2\theta} )$. All the metric coefficients are functions 
only of $\r$ and $t$.  The dynamical quantities in the $3+1$ initial
value formulation are the intrinsic metric $\gamma_{ab}$, and the
extrinsic curvature $K_{ab}$, of the slices. In spherical symmetry,
the extrinsic curvature has only two independent components: $K^{i}_{j}
= \text{diag}(H_A,H_B,H_B)$. 

The spatial projections of Einstein's equations lead to the 
evolution equations for the spatial tensors $\gamma_{ij}$ and $K^{i}_{j}
= \gamma^{ik} K_{kj}$,
\begin{eqnarray}
A_{,t} \, = \,
&& - \alpha H_{A} + e^{-2A} ( \beta_{,\r} - \beta A_{,\r}) ,
\label{eq:adot}\\
B_{,t} \, =  \, 
&& - \alpha H_{B} + e^{-2A} \beta \left(\frac{1}{\r} + B_{,\r} \right) ,
\label{eq:bdot}\\
H_{A,t} =
&& e^{-2A} [  \alpha R_{\r\r} + \beta H_{A,\r} - \alpha_{,\r\r} 
              + A_{,\r} \alpha_{,\r} ]
   + \alpha H_A K  -\kappa [e^{-2A} (\Phi_{,\r})^2]  ,
\label{eq:hadot}\\
H_{B,t} =
&&  e^{-2A} \left[  \alpha \left(\frac{e^{2A -2B}}{\r^2}\right)  R_{\theta\theta} 
             + \beta H_{B,\r} 
             - \alpha_{,\r} \left(\frac{1}{\r} + B_{,\r} \right)\right]
   + \alpha H_B K  , 
\label{eq:hbdot}
\end{eqnarray}
where  $K$ denotes the trace $K^{i}_{i}$ of the extrinsic curvature.

The time-time and time-space projections of the  Einstein's equations yield
respectively the Hamiltonian and momentum constraints
\begin{equation}
  \frac{1}{2}R_{\r \r} 
+ \frac{e^{2A-2B}}{\r^2} R_{\theta\theta} + e^{2A} H_B (H_B + 2 H_A)
= \frac{1}{2}\kappa  \left[ (\Phi_{,\r})^2 +  e^{2 A} \Pi^2 \right] 
\label{eq:H}
\end{equation}
and
\begin{equation}
H_{B,\r} +  \left(\frac{1}{\r} +  B_{,\r} \right) (H_B - H_A)
= \frac{1}{2}\kappa \Pi \Phi_{,\r} ~,
\label{eq:M}
\end{equation}
where $R_{\r\r}$ and $R_{\theta\theta}$ are the 3-Ricci
components given by
\begin{eqnarray}
\frac{1}{2}R_{\r\r} = &&  
- B_{,\r\r} + \frac{1}{\r}(A_{,\r} 
- 2 B_{,\r}) + B_{,\r} (A_{,\r} - B_{,\r}) ,   
\label{eq:Ree} \\
 \frac{e^{2A-2B}}{\r^2} R_{\theta\theta} = &&  
- B_{,\r\r} + \frac{1}{\r}(A_{,\r} 
- 4 B_{,\r}) + B_{,\r} (A_{,\r} - 2 B_{,\r})
+ \frac{1}{\r^2} (e^{2A-2B} - 1) .
\label{eq:Rtt}
\end{eqnarray}
Finally, the 3$+$1, first order in time, form of the Klein-Gordon
equation (\ref{eq:klein}) is
\begin{eqnarray}
\Phi_{,t} = && \alpha\,\Pi +  e^{-2A} \beta \Phi_{,\r} , 
\label{eq:kgs1} \\
\Pi_{,t}  = && 
 e^{-2A} \left\{  \alpha \left[ 
   \frac{1}{\r^2} \left(\r^2 \Phi_{,\r} \right)_{,\r}
+  (2B_{,\r} - A_{,\r}) \Phi_{,\r} 
   + e^{2A} K \Pi \right]
   +  \alpha_{,\r} \Phi_{,\r} + \beta \Pi_{,\r} \right\}.  
\label{eq:kgs2}
\end{eqnarray}

The geometry evolution equations (\ref{eq:adot})-(\ref{eq:hbdot})
together with the matter evolution equations
(\ref{eq:kgs1})-(\ref{eq:kgs2}) constitute an initial value problem
for the quantities $A, B, H_{A}, H_{B}, \Phi, \Pi$. The initial
data for the IVP must satisfy the constraints
(\ref{eq:H})-(\ref{eq:M}). Obtaining the development of the initial
data requires, furthermore, the specification of the gauge functions
$\alpha$ and $\beta$.  Ideally, the time integration of the 
IVP defined above should put no restrictions on either the form of the
metric or the gauge functions. Yet the use of spherical coordinates,
mandated by the Killing symmetries (and also
computationally efficient), significantly limits the freedom to choose
the space-time slicing. For example, the simplest possible gauge, the
geodesic or synchronous gauge ($\alpha = 1, \beta = 0$),
leads to a coupled system of
equations for the geometric variables $A,B,H_{A},H_{B}$.
Formulating this problem as set of second order equations for $A,B$
reveals a dynamical structure that is not that of a wave
system. In practice, this leads to considerable difficulties in 
preserving the
appropriate regularity of the geometry near the origin.
Such regularity is analytically ensured by cancelations of
$1/r$ terms both explicitly in the right-hand side of the evolution equations
and implicitly by the enforcement of the constraints.

A gauge choice that overcomes the regularity problems at the origin is
the radial gauge, in which $\r$ is chosen such that the area of each
$dt = d\r = 0$ sphere is equal to $4 \pi
\r^2$, which leads to the condition $B=0$.  The integration procedure
outlined below also assumes a vanishing shift condition, but 
can be generalized for arbitrary shift.

The line element (\ref{eq:met}) with $B = \beta = 0$ becomes
\begin{equation}
   ds^2 = -\alpha^2 dt^2  +  e^{2A} \,d\r^2 
   + \r^2  (d\theta^2 +\sin^2\theta d\phi^2).
\label{eq:cmetric}
\end{equation}

As an immediate consequence of $B = \beta = 0$, equation (\ref{eq:bdot}) 
implies that $H_{B}$ also vanishes, and
equation (\ref{eq:hbdot}) for $H_{B,t}$ gives a condition
on the lapse  function,
\begin{equation}
  \alpha_{,\r}  -  \left[ A_{,\r} + \frac{1}{\r}(e^{2A}-1)\right] \,  
  \alpha  = 0 ~.
\label{eq:lapse_cond}
\end{equation}
Furthermore, equations (\ref{eq:adot}) and 
(\ref{eq:hadot}) reduce to
\begin{eqnarray}
A_{,t} \, = \, && - \alpha H_{A}  ~,
\label{eq:new_adot} \\
H_{A,t} = && e^{-2A} \left[ \frac{2}{\r}  \alpha A_{,\r}
                     - \alpha_{,\r\r} + A_{,\r} \alpha_{,\r} \right]
   + \alpha H_{A}^2  -\kappa e^{-2A} (\Phi_{,\r})^2 ~.
\label{eq:new_hadot}
\end{eqnarray}
With the conditions $B = H_B = 0$, 
the Hamiltonian and momentum constraints are now decoupled equations
relating the metric function  $A$ and its associated extrinsic curvature $H_A$
to the scalar field energy density and current respectively:
\begin{equation}
A_{,\r} + \frac{1}{2\r}(e^{2A} - 1) = \frac{1}{2} \kappa \r 
    \left[ (\Phi_{,\r})^2  +  e^{2A}  \Pi^2 \right]  ~,
\label{eq:new_hc}
\end{equation}
\begin{equation}
H_A = - \frac{1}{2} \kappa \r  \Pi \Phi_{,\r} ~.
\label{eq:new_mom}
\end{equation}
Finally, the scalar field equations, in this gauge, take the form
\begin{eqnarray}
\Phi_{,t} = && \alpha\,\Pi  ~, 
\label{eq:sw1} \\
\Pi_{,t}  = && 
 \alpha  e^{-A} 
   \left[  \frac{1}{\r^2} \left(\r^2 \Phi_{,\r} \right)_{,\r}
           - \frac{A_{,\r}}{2} \Phi_{,\r}  + e^{A} H_{A} \Pi  \right]
   + e^{-A} \alpha_{,\r} \Phi_{,\r}  ~.  
\label{eq:sw2}
\end{eqnarray}

In summary, the lapse function is determined by the condition
(\ref{eq:lapse_cond}). The remaining gravitational variables ($A$ and
$H_A$) can be constructed from a subset of the evolution equations
(\ref{eq:new_adot}-\ref{eq:new_hadot}) and constraint equations
(\ref{eq:new_hc}-\ref{eq:new_mom}). Different integration schemes can
be designed depending on which two equations are chosen from
(\ref{eq:new_adot}-\ref{eq:new_mom}) to solve for $A$ and $H_A$. For
instance, a free, unconstrained evolution consists of using equations
(\ref{eq:new_adot}) and (\ref{eq:new_hadot}).  On the other hand, a
fully constrained evolution would require solving the Hamiltonian
constraint (\ref{eq:new_hc}), with the extrinsic curvature $H_{A}$
computed from the source terms using (\ref{eq:new_mom}). 
Alternatively, a mixed scheme can be
followed, with $H_{A}$ again given by (\ref{eq:new_mom}), while the
metric variable is updated using (\ref{eq:new_adot}), subject again to
the lapse condition. Still other alternative schemes
involving combinations of (\ref{eq:new_adot}-\ref{eq:new_mom}) 
are possible. A fully constrained evolution is adopted here
since it has been used in earlier accurate calculations of scalar wave
collapse \cite{Choptuik-print} and facilitates enforcing a regular
metric boundary condition at the origin.

In order to complete the IVP, specific boundary conditions for the
scalar field variables $\Phi, \Pi$ at both ends of the integration
domain must be prescribed, along with integration constants for the
hypersurface equations (\ref{eq:new_hc}) and (\ref{eq:lapse_cond}).
The scalar field variables must be finite at $\r=0$, and thus the
appropriate boundary conditions at this point are $\r \Phi = \r \Pi =
0$. The outer boundary conditions for these variables are well
understood only in the limit $\r \rightarrow
\infty$, where rigorous  outgoing wave conditions exist.
Imposing a boundary condition at any finite $\r$ involves a certain
{\em physical} approximation. Achieving a complete solution of the
IVP without such additional assumptions is indeed the main focus of
the CCM program, and the details are given in section
\ref{sec:ccm}.

The prescription of the integration constants for the hypersurface
equations at some point $\r_{0}$ relates the labeling of the time and
radial coordinates with the proper time and radial distance
measurements of a privileged observer. Two obvious choices are geodesic
observers, either at the center of symmetry, or at the outer boundary.
While the choice does not have any consequences on physical
observables, it is more natural for our integration procedure to assume
a geodetic observer at the center of symmetry, and hence to require
that $A = \alpha = 0$ there.  In fact, this choice is well suited for
the study of critical collapse as well, where the self-similar critical
solution \cite{Choptuik-PRL} manifests itself in terms of the proper
time of an observer at the origin.

\section{Exterior Domain: Characteristic Evolution}
\label{sec:null}

For the exterior characteristic evolution, a family of null
hypersurfaces $u=const$ is introduced, emanating along the outward
normals to the cross-sections of the matching world-tube and labeled by
$u=t$.  The outgoing null rays are parametrized by an area coordinate
$r$, with $r=R$ at the matching worldtube. The coordinate $x=r/(R+r)$
is introduced for purposes of compactification, so that null infinity
is located at $x=1$. (In the numerical simulations to be presented
later, we set the scale so that $R=1$ and consequently $x=1/2$ at the
matching worldtube). In the null coordinate system, the line element in
the exterior region has the form
\begin{equation}     
    ds^2=-e^{2\lambda}du({V \over r}du+2dr) 
    + r^2 (d\theta^2 +\sin^2\theta d\phi^2), 
\label{eq:nmetric}
\end{equation} 
where the metric functions $\lambda$ and $V$ depend only on $u$ and
$r$.

The hypersurface equations for $\lambda$ and $V$ are
\begin{equation}
    \lambda_{,r} = 2\pi r(\Phi_{,r})^2    
\label{eq:lambda}
\end{equation}
and
\begin{equation}
     V_{,r}=e^{2\lambda}.             
\label{eq:V}
\end{equation}
The scalar wave equation in the characteristic region takes the form
\begin{equation}
     2 (r \Phi_{,u})_{,r} = \frac{1}{r} (V \Phi_{,r} )_{,r}
\label{eq:SWE}
\end{equation}
In terms of the intrinsic metric of the $(u,r)$ sub-manifold,
\begin{equation}
    \eta_{\alpha\beta}dx^{\alpha} dx^{\beta} 
    =-e^{2\lambda}du({V \over r}du+2dr), 
\end{equation}
where the Greek indices take the values $(0,1)$, this reduces to
\begin{equation}
     \Box ^{(2)}g= {e^{-2\lambda} g\over r} ({V\over r})_{,r},
\label{eq:hatwave}
\end{equation} 
where  $g=r\Phi$ and $\Box ^{(2)}$ is the D'Alembertian associated with 
$\eta_{\alpha \beta}$. 

The integration of the system
(\ref{eq:lambda})-(\ref{eq:SWE}) proceeds with the
specification of initial data $\Phi(u_0,r)$ for $r>R$ on the initial
null cone $u_{0}$. The hypersurface equations are integrated radially
and furnish compatible geometric functions $\lambda$ and $V$. This in
turn allows the time integration of the scalar field equation, which
provides new data in the neighborhood $u_{0} + du$ and completes the
integration cycle.  The extra information needed to evolve the
space-time is the characteristic initial data $\Phi(u,R)$ at the
inner boundary, as well as the boundary values
$\lambda(u,R)$ and $V(u,R)$.  A characteristic integration algorithm for
(\ref{eq:hatwave}) may be based upon the null parallelogram made up of
incoming and outgoing radial characteristics \cite{jcomp,Null-cone}. 

\section{Cauchy-Characteristic Matching}
\label{sec:ccm}

The match is performed across a three-dimensional, time-like
hypersurface $\Sigma$ (world-tube), which divides the four-dimensional
space-time into two disjoint sub-manifolds $M^+$ (Characteristic
exterior) and $M^-$ (Cauchy interior).  Each sub-manifold $M^\pm$ is
endowed with a metric $g_{ab}^\pm$ that induces a unique intrinsic
geometry at $\Sigma$.  Independent coordinate charts $\lbrace x^a
\rbrace^\pm$ are introduced in $M^\pm$. Let $s^a$ be the space-like,
unit normal to $\Sigma$ directed from $M^-$ to $M^+$.  The metric
intrinsic to $\Sigma$ is then given by

\begin{equation}
h_{ab} = g_{ab} - s_a s_b,
\end{equation}
where $h^a_b$ is the projection operator into the subspace $\Sigma$.
The second fundamental form, or extrinsic curvature  of $\Sigma$ is defined by
\begin{equation}
\Theta_{ab} = h_a^ch_b^d \nabla_{(c} s_{d)}.
\end{equation}
If $v^a$ is the time-like, unit tangent to $\Sigma$, the metric $h_{ab}$
has the further decomposition
\begin{equation}
h_{ab}=  -v_a v_b + R^2 q_{ab},
\end{equation}
where $q_{ab}$ is the unit sphere metric.
The metric tensors $h_{ab}, q_{ab}$ and the vectors $s^a, v^a$ satisfy
the following orthogonality conditions: $s^a h_{ab} = s^a q_{ab} = v^a
q_{ab} = s^a v_a = 0$.

The metric continuity requirement can be recast as conditions on the
induced norm of the tangent vector $v^a$ on $\Sigma$,
\begin{equation}
[v^a v^b g_{ab} ]^- = [v^a v^b g_{ab} ]^+
\label{eq:m_cont1}
\end{equation}
and the surface-area radius,
\begin{equation}
[q^{ab} g_{ab} ]^- = [q^{ab} g_{ab} ]^+.
\label{eq:m_cont2}
\end{equation}
In addition to the continuity of the metric across $\Sigma$, in order
to prevent sheet discontinuities (singular hypersurfaces), 
the following conditions on the extrinsic curvature must be imposed:
\begin{equation}
[v^a v^b \Theta_{ab} ]^- = [v^a v^b \Theta_{ab} ]^+
\label{eq:k_cont1}
\end{equation}
and 
\begin{equation}
[q^{ab} \Theta_{ab} ]^- = [q^{ab} \Theta_{ab} ]^+.
\label{eq:k_cont2}
\end{equation}
Equations (\ref{eq:m_cont1})-(\ref{eq:k_cont2}) should be understood 
as a limit process. Namely for any tensor $A_{ab}$, the continuity
condition $[A_{ab}]^- = [A_{ab}]^+$ implies
\begin{equation}
\lim_{Q\rightarrow P^-} A_{ab}^-(Q) = \lim_{O\rightarrow P^+} A_{ab}^+(O),
\label{eq:l_cont}
\end{equation}
where $P^\pm \in \Sigma^\pm$ and $Q \rightarrow P^-, O \rightarrow
P^+$, through $M^-,M^+$, respectively. The continuity condition
(\ref{eq:l_cont}) has meaning only if there exists a mapping ${\cal
M}: \Sigma^- \rightarrow \Sigma^+$ which transforms tensor
components between the two coordinate systems.
 
In the spherically symmetric case under consideration, the coordinate
systems $\lbrace t, \r, \theta, \phi \rbrace$ in $M^-$ and $\lbrace u,
r, \theta, \phi \rbrace$ in $M^+$ are introduced.  Because of the
symmetry, the same coordinates $\theta$ and $\phi$ are used everywhere
throughout $M^\pm$. This is not the case with the radial coordinate.  In
general, $\r$ and $r$ are not the same.  The line elements in $M^\pm$
are given by (\ref{eq:cmetric}) and (\ref{eq:nmetric}), respectively.
The tangent vector to $\Sigma$ in these coordinates is
\begin{equation}
v^a = \frac{dx^a}{d\tau} \equiv \dot x^a= \cases{
(\dot t, \dot \r, 0 , 0) & in $M^-$\cr
\noalign{\vskip2pt}
(\dot u, \dot r, 0 , 0) & in $M^+$.\cr}
\end{equation}
In this section, a dot denotes derivative with respect to the proper
time $\tau$ along the world tube $\Sigma$. 
That is $\dot{f} \equiv \frac{d f}{d\tau} = v^a\partial_a f$.
From $u^a s_a = 0$ and $s^as_a = 1$, it follows that
\begin{equation}
s_a = \cases{
\alpha e^{A}\,(-\dot  \r , \dot t , 0 , 0)  & in $M^-$\cr
\noalign{\vskip2pt}
e^{2\lambda}\,(-\dot r, \dot u , 0 , 0)  & in $M^+$,\cr}
\end{equation}
and
\begin{equation}
s^a = \cases{ \alpha^{-1} e^{A} (
 \dot \r, \alpha^2 e^{-A} \dot t , 0 , 0)  & in $M^-$\cr
\noalign{\vskip2pt}
(-\dot u, \dot r + \dot u V/r, 0 , 0)  & in $M^+$,\cr}
\label{s_up}
\end{equation}

The continuity conditions on the metric, 
(\ref{eq:m_cont1}) and (\ref{eq:m_cont2}), can now be rewritten  as
\begin{equation}
-\alpha^2 \dot t^2 + e^{2A} \dot \r^2 = 
-e^{2\lambda} (\dot u^2 V/r + 2 \dot u \dot r)
\label{gcont}
\end{equation}
and
\begin{equation}
\r = r \equiv R,
\end{equation}
respectively, where $R$ is the surface-area radius of the matching
surface. $R$ is taken to be constant on $\Sigma$, hence the matching
surface is invariant under the orbits of the spherical symmetry. 

An additional assumption adopted here is  that the matching surface
does not move in coordinate space ($\dot r = \dot \r = 0$) and
that the coordinate time in $M^+$ and $M^-$ agree in $\Sigma$ ($u =
t$).  The metric continuity condition (\ref{gcont}) then reduces to
\begin{equation}
\alpha^2  =  e^{2 \lambda} \frac{V}{R},
\label{match1}
\end{equation}
with all functions evaluated at the matching surface.

Similarly, the continuity conditions on the extrinsic curvature,
(\ref{eq:k_cont1}) and (\ref{eq:k_cont2}), yield respectively
\begin{equation}
[q^{ab}\nabla_a s_b]^- = [q^{ab}\nabla_a s_b]^+
\label{k2_cont1}
\end{equation}
and
\begin{equation}
[s^a a_a]^- = [s^aa_a]^+,
\label{k2_cont2}
\end{equation}
where $a^a = u^b \nabla_b u^a$ is the world tube ``acceleration."
In spherical symmetry, condition (\ref{k2_cont1}) reduces to
\begin{equation}
[s^{a} \nabla_a R]^{+} = [s^{a} \nabla_{a} R]^{-} ,
\end{equation}
which can be rewritten using (\ref{s_up}) explicitly in terms of the metric
functions as
\begin{equation}
\frac{\alpha}{e^A} = \frac{V}{R},
\label{match2}
\end{equation}
at the matching surface $\Sigma$.

The condition (\ref{k2_cont2}) yields an equation for
$\ddot{t}$ from which the coordinate times, $t$ or $u$, at $\Sigma$
can be solved in terms of the proper time $\tau$. Alternatively,
$t(\tau)$ and $u(\tau)$ can be obtained from the normalization condition
of $v^a$ of $s^a$.

The matching of the scalar field variables across the two coordinate
domains must ensure that neither the field values nor the field derivatives
exhibit jump discontinuities on the interface. That is,
\begin{equation}
 [\Phi]^{-} = [\Phi]^{+}
\end{equation}
and
\begin{equation}
 [l^{a} \nabla_{a} \Phi]^{-} = [l^{a} \nabla_{a} \Phi]^{+}
\end{equation}
for any vector $l^{a}$.
It is convenient to choose $l^{a} = s^{a} + v^{a}$, which is an outgoing
null vector at $\Sigma$. In this case, the continuity equation
at the matching surface becomes
\begin{equation}
\frac{\partial \Phi}{\partial t} + \alpha e^{- A} 
\frac{\partial \Phi}{\partial \r} =
\frac{V}{R} \frac{\partial \Phi}{\partial r}.
\label{eq:cont1}
\end{equation}
In the limit $R \rightarrow \infty$, for an asymptotically flat
space-time with no incoming radiation, equation~(\ref{eq:cont1})
reduces to the familiar Sommerfeld condition if the right-hand side of
(\ref{eq:cont1}) is set to zero. However, unlike the Sommerfeld
condition which is only valid asymptotically, equation~(\ref{eq:cont1})
is an {\it exact} relation, valid at any distance.

For the momentum variable $\Pi$, the procedure is different, to avoid
imposing continuity on higher derivatives at $\Sigma$.
Starting from the definition $\Pi = v^{a} \nabla_{a} \Phi$, it is
rewritten as
\begin{equation}
\Pi + s^{a} \nabla_{a} \Phi = l^a \nabla_{a} \Phi
\end{equation}
which leads to the condition
\begin{equation}
e^{A} \Pi + \frac{\partial \Phi}{\partial \r} = 
\frac{\partial\Phi}{\partial r}
\label{eq:cont2}
\end{equation}

The system of equations (\ref{match1}),(\ref{match2}),(\ref{eq:cont1})
and (\ref{eq:cont2}) completes the specification of the matching 
interface.

\section{Tests and Results}
\label{sec:tests}

\subsection{Stability and accuracy tests}

The discretization algorithms in both domains, as well as the
implementation of the continuity conditions are all constructed by
replacing derivatives by second order accurate, centered finite
differences. The matching surface lies at a fixed coordinate location,
which is a grid point of both the $\r$ and $r$ coordinate grids. This
simple scheme leads to long term numerical stability, which is not an
automatic feature of a matching algorithm \cite{3Dccm}. Here, long term
stability is defined as the bound evolution of initial data over time
periods large compared to the light crossing time of the inner
computational domain. This stability requirement is stronger than
classical Von-Neumann stability, which requires bound local propagation
of linearized modes.  Such stability is becoming more and more
important in numerical relativity as the desired integration times
become longer. The CCM code developed in this work exhibited stability
for any practical evolution time.

In Fig.~\ref{fig:evolution} a typical matched evolution is shown for
initial data with a relatively small $M/R$ ratio ($0.08$).
The functional dependence of the initial data is given in this case by a
Gaussian,
\begin{equation}
 r \Phi = \lambda e^{-((r - r_{c})/\sigma)^2}
\label{eq:gaussian}
\end{equation}
with $\lambda = 0.0225$, $r_{c} = 2.0$ and $\sigma = 0.1$. 
The left column shows snapshots of the evolution in the Cauchy domain,
with the radial coordinate $\r$ running from the origin $\r = 0$ to
the matching radius $\r = 1$. The right column shows the corresponding
null evolution, with the compactified radial coordinate $x$ running
from the matching radius $x=0.5$ to null infinity at $x=1$. The field
variables illustrated in all snapshots are $\r \Phi$ and $r \Phi$, with
the scale being constant for plots in the same row.

The first row includes a few snapshots that follow an imploding pulse
as it crosses the matching surface and propagates into the Cauchy
domain. Note that at this instance, an inaccurate matching scheme
would create back-reflection which would immediately register at null
infinity.  The apparent widening of the pulse as it enters the Cauchy
region is due to the different radial functions used in the two
coordinate systems as well as the doubling of the local propagation
speed in the Cauchy sector.  The second row demonstrates the
propagation and reflection of the (marginally sub-critical) pulse off
the origin. Strong non-linear distortions occur there, while the
leading part of the pulse is already crossing the matching surface
and radiating to null infinity.  In the third row, the peak of the
pulse is  propagating outward across the matching surface; and the
solution finally decays in the fourth row, as the trailing parts of
the pulse cross $\Sigma$. The very slight curvature of the pulse in the null
region, as the peak amplitude crosses the matching surface, indicates
a small amount of backscattering occurring at this time.

Second order convergence of all computed quantities to a
limiting value is readily verified, e.g., by monitoring the final
field configuration for a sequence of successively refined grids. A
more physically intuitive test, the conservation of the total energy of
the system, is a powerful probe into how well the discretization of
the Einstein equations preserves the additional differential structure
encoded in the Bianchi identities.  A test of the absolute convergence
of the energy residual $\Delta M$ is performed for that purpose.  The
energy residual between two time levels $u_1, u_2$ is defined as
\begin{equation}
\Delta M(u_1,u_2) = M(u_1) - M(u_2) + \int_{u_1}^{u_2} P(u)~du
\end{equation}
where $M(u)$ is the Bondi mass content of the space-time slice at time $u$
while $P(u)$ is the power flow at infinity.  
$\Delta M$ is identically zero in the continuum limit and thus
must converge to zero appropriately as a function of the 
discretization length-scale $\Delta$. 

The explicit forms of the Bondi mass $M$ and the radiated power
$P$  in terms of metric quantities are given by 
\begin{equation}
M(u) = \left. \frac{1}{2} e^{-2H} r^2 (\frac{V}{r})_{,r} \right|_{\infty}
\end{equation}
\begin{equation}
P(u) = - 4 \pi e^{- 2 H} Q_{,u}^2 \label{eq:P}
\end{equation}
where 
$H(u) = \lim_{r\rightarrow \infty} \lambda(u,r)$
and 
$Q(u)= \lim_{r\rightarrow \infty} r \Phi(u,r)$.

In Fig.~\ref{fig:energy_conv} the convergence of the energy
residual is demonstrated. A sequence of approximate solutions with progressively
finer resolution are obtained. The initial data (\ref{eq:gaussian}) are
prescribed in the null sector, while the Cauchy sector is taken to be
flat initially.  The Bondi mass is computed at the initial Bondi time
$u_1 = 0$ and at a fixed final Bondi time $u_2 = 4.0$, while the power
integral is accumulated (to second order accuracy) at each integration
step. The second order convergence of the energy residual, consistent
with the second order discretization of the component algorithms and
the matching interface, is evidence of a successful matching of the two
evolution schemes. The computational error in the mass and, most
importantly, in the radiated power is directly controlled by the grid
spacing $\Delta$, and in fact diminishes as $\Delta^2$.

\subsection{Computing a space-time with two alternative foliations}

A comparison of the numerical solution obtained by two, considerably
different foliations of our model spacetime is illustrated next. First,
a calibrated characteristic code \cite{Asymptotics} is used to obtain a
global evolution of initial data corresponding to an incoming wave with
support outside a radius $x_1 = r_1 / (1 + r_1)$
(Fig.~\ref{fig:null}).  This one-dimensional characteristic evolution
will supply the waveform of the outgoing radiation coming back out to
future null infinity.  Since the initial data have compact support
outside $x_{1}$, the space-time portion delimited by the initial time
($u=0$) surface, the origin world-line $x=0$ and the incoming null cone
$C$ (beginning at $x_1$, will be flat.

Next, the CCM code, with a matching radius at $x=1/2<x_1$, is used to
evolve the same characteristic data, along with flat-space Cauchy data
on the initial time ($t=0$) surface (Fig.~\ref{fig:ccm}).  The incoming
pulse enters the Cauchy region across the matching surface in the
inward direction, then gets Cauchy evolved until it leaves the matching
surface in the outward direction and ends up at future null infinity.
This test compares the waveforms at null infinity produced by a global
null code and by a CCM code.  Theoretically, the general covariance of
the equations guarantees that the output should be identical.  In
practice, this test checks a combined algorithm (CCM) against a well
calibrated scheme, i.e., the one produced by the global null
algorithm.

The waveforms are compared at null infinity, as would be measured by
asymptotic (Bondi) observers. However, the time coordinates are 
considerably different for the two evolution schemes. In the CCM
approach, the central time  parametrizes the space-like
foliation with the proper time of an observer at the center of
symmetry. This parametrization subsequently also labels the outgoing
null cone foliation of the outer region of the space-time, with the
synchronization performed at the matching radius.  In contrast, the
time parameter of the null foliation follows the proper time of the
central observer, directly labeling the outgoing null cones 
emanating from the center of symmetry. Before comparing the two
signals $(g_{CCM}(t) , g_{N}(u))$ they must be reparametrized 
according to the asymptotic Bondi time.

In a Bondi frame, geodesic observers would measure an asymptotically
Lorentzian line element, which  in the characteristic coordinate system
is
\begin{equation}
ds^2 = - d\tilde{u}^2 - 2 d\tilde{u} dr 
+ r^2 (d\theta^2 +\sin^2\theta d\phi^2).
\end{equation}
The Bondi time $\tilde{u}$ is related to the coordinate time $u$ of a
general  null cone foliation by the factor
\begin{equation}
\frac{d\tilde{u}}{du} = e^{2 H(u)},
\end{equation}
defined following equation (\ref{eq:P}).

In Fig.~\ref{fig:comp} the signals at null infinity 
and their difference are shown as a function of Bondi time.
The initial (characteristic) data are
\begin{equation}
  \Phi  = \Lambda  r^2  e^{ - (r - r_{c})^2 / \sigma^2} \sin{k r}
\label{eq:mod_gaussian}
\end{equation}
with the parameters for this plot being $\Lambda = 6 \times 10^{-4}$,
$r_{c} = 3.0$, $\sigma = 0.6$ and $k=10$. This value of $\Lambda$ is
just below the threshold of black hole formation (which occurs at
about $\Lambda = 6.125 \times 10^{-4}$) and leads to the strong
distortion of the signal in the second half of the pulse.  The signals
obtained with the two codes overlayed as functions of Bondi time show
little difference to graphical accuracy, a manifestation of physical
covariance and algorithmic compliance.  The grid sizes used for this
run where $500 + 500$ points for the CCM code, and $1000$ for the null
code. The relative difference between the two signals for those
resolutions is at the level of 0.1\%.  The maximum absolute value
difference between the two signals over the total integration time
provides a strong and physically interesting norm. This norm converges
to zero with a measured rate of $1.99$, consistent with the
anticipated second order convergence.

The investigation is now extended to strong field phenomena, with the
study of initial data that end up in the formation of a black hole. In
the Cauchy region, black hole formation is signaled when
$e^{2A}\rightarrow\infty$ and the function $2 m/r= 1 -
e^{-2A}\rightarrow 1$ at some radius
$R_{BH}$. The mass of the black hole is then $M_{BH}= 2
R_{BH}$. Alternatively, the behavior of the lapse function, or the
scalar field values themselves can be used for similar estimates.  In
comparison, the formation of the black hole in the null region is
signaled by the infinite red-shift between Bondi time and coordinate
time for all $r > R_{BH}$, and the resulting decay of further
radiative loss to infinity.  The final value of the Bondi mass is
then equal to $M_{BH}$.

In practice, the evolution of the system cannot proceed accurately long
after the black hole has formed. The use of a surface area coordinate
$r$ leads to a coordinate singularity inside the horizon \cite{Christo}
and the premise of the validity of the finite difference approximation
breaks down.  This is reflected, for example, in the breakdown of the
convergence properties of the algorithm. Nevertheless, the matching
procedure is sufficiently accurate to allow an investigation of black
hole formation quite close to the matching surface.  (The mass of the
black hole, and consequently the $M_{BH}/R$ ratio is controlled
directly by the parametrization of the initial data.)

Fig.~\ref{fig:bh-evol-ccm} shows the evolution of a strong Gaussian
pulse, leading to a black hole that has a mass only slightly smaller
than the mass scale set by the matching radius ($R=1$).  For
demonstration purposes, the radial coordinate $\r$ is also
compactified here, and the field functions are shown as if they were defined
on a single grid. The highly non-linear pulse creates strongly
back-scattered radiation as it propagates inwards, but not enough
to avoid a black hole collapse.  The discontinuity of the field
derivatives at the location of the horizon is prominent in the
late time snapshot.

Fig.~\ref{fig:distort} displays the effects of black hole formation on
the time development of the waveform and of the Bondi mass. The initial
data for those runs are given by (\ref{eq:mod_gaussian}) with the
amplitude $\Lambda$ increased progressively from $5.50 \times 10^{-4}$
to $6.25 \times 10^{-4}$ in three steps.  Up until Bondi time of about
7.0, an increase in the amplitude of the initial data leads to a near
linear increase of the initial mass (lower plot) and to a corresponding
red-shifting of the waveform. Afterwards, there is a bifurcation point,
at which data that lead to a black hole space-time produce a signal
severely distorted with respect to that of non-collapsing data. The
Bondi mass either drops to zero for an asymptotically Minkowskian
space-time, or coasts to a constant value that represents the mass of
the newly formed black hole.

\section{Conclusion}
\label{sec:conclusions}

The model problem of the self-gravitating scalar-field provides a
convenient framework for the study of the CCM research program in a
controlled situation. It has been demonstrated here that the matched
evolution is essentially transparent to the presence of the
interface. The targets of (1) clearly identified radiative quantities and
(2) physically accurate boundary conditions are achieved in this case with
minimal computational and developmental effort. Direct application of
the continuity conditions at the matching interface leads to a
stable and accurate mixed evolution algorithm for all relevant $M/R$
ratios. Although idealized, this model problem captures many essential
physical aspects of more generic asymptotically flat space-times. It is
expected that the geometrical approach and the algorithmic methodology
initiated here are applicable and useful in more realistic problems, and
work is currently under way exploring this possibility.

\section{Acknowledgments}

We thank M. Choptuik and E. Seidel for helpful discussions of origin
singularities in spherically symmetric systems.  This work was
supported by the Binary Black Hole Grand Challenge Alliance, NSF
PHY/ASC 9318152 (ARPA supplemented) and by NSF PHY 9510895 to the
University of Pittsburgh. P.L was supported in part by NSF Young
Investigator award PHY 93-57219 and NSF grant PHY 93-09834.  Computer
time for this project has been provided by the Pittsburgh
Supercomputing Center under PHY860023P to J.W.

\bibliographystyle{unsrt}

\newpage 

\begin{figure}
\caption{CCM evolution. The left column depicts evolution in the
Cauchy domain, the right column shows the corresponding null
evolution.  The field variables illustrated are $\r \Phi$ and $r \Phi$
respectively. For each row, the left hand scale gives the amplitude in
the two domains.}
\label{fig:evolution}
\end{figure}

\begin{figure}
\caption{Convergence test for the energy residual $\Delta M$.  The
grid size $\Delta \r$ refers to the Cauchy sector. The sequence of
successive higher resolutions maintains a fixed ratio of the null
sector grid size ($\Delta r$) to that of the Cauchy sector grid size
($\Delta \r$).  The convergence rate for the demonstrated sequence of
grid resolutions is 1.89.}
\label{fig:energy_conv}
\end{figure}

\begin{figure}
\caption{The foliation of an asymptotically flat spherically symmetric
space-time by outgoing null cones emanating from the origin.
The initial data have compact support outside $x_{1}$ so that the
inner region of the space-time (bounded above by $C$) is flat.
The vertical and horizontal dotted lines give the location of the
matching radius and the initial data surface of the comparison
CCM run. }
\vspace{1cm}
\label{fig:null}
\end{figure}

\begin{figure}
\caption{The foliation of an asymptotically flat spherically symmetric
space-time with a combination of space-like and null hypersurfaces.
Initial data for the comparison with the globally null code are given
outside the matching surface. Flat Cauchy data in the interior
complete the specification of a physical system.}
\vspace{1cm}
\label{fig:ccm}
\end{figure}

\begin{figure}
\caption{Comparison of null and CCM evolutions in the case of
sub-critical initial data. The upper plot demonstrates the nearly
identical signal at null infinity. The solid line shows the signal
output of the CCM code, while the circles indicate
the corresponding null code output at the same Bondi time.
The lower plot highlights the difference in the signals. 
The signal difference for a CCM grid of 500+500 points and a Null grid of 1000
points, is illustrated by the circle line. The diamond line shows the 
second order reduction of the difference when the resolution of 
all grids is doubled.}
\label{fig:comp}
\end{figure}

\begin{figure}
\caption{CCM evolution and black hole formation. The evolution of
sufficiently strong Gaussian initial data leads to the formation of an
apparent horizon, in this case at about $\r = 0.8$ ($x=0.45$).  This
is just marginally inside the matching radius at $\r = 1$
($x=0.5$). The incoming Gaussian pulse (right side of the plot)
propagates inwards (to the left), crosses the matching surface and 
collapses, forming a cusp-like profile.}
\label{fig:bh-evol-ccm}
\end{figure}

\begin{figure}
\caption{The distortion of the signal at infinity for a space-time
that develops a black hole. The upper plot shows overlapping
signals with a progression of initial amplitudes. The dotted
signal corresponds to black hole formation. The lower plot
shows the evolution of the total Bondi mass.}
\label{fig:distort}
\end{figure}

\end{document}